\documentclass[reprint,amsmath,amssymb,aps,prd,noeprint,nolongbibliography,superscriptaddress]{revtex4-2}
\usepackage{graphicx,xcolor,verbatim,makecell,multirow,float,mathtools,slashed} 
\usepackage[colorlinks,citecolor=blue,urlcolor=blue,linkcolor=blue]{hyperref}
\usepackage[normalem]{ulem}
\makeatletter
\def\UTFviii@defined#1{\ifx#1\relax!!FIXME!!\else\expandafter#1\fi} 
\makeatother
\definecolor{nicegreen}{rgb}{0.,0.5,0.}
\begin{document}
\title{Muon collider probes of the gluonic quartic gauge couplings}
\author{Yu-Chen Guo}
\email{ycguo@lnnu.edu.cn}
\affiliation{Center for Theoretical and Experimental High Energy Physics, Department of Physics,
Liaoning Normal University, Dalian 116029, China}
\affiliation{Pittsburgh Particle Physics, Astrophysics, and Cosmology Center, \\
Department of Physics and Astronomy, University of Pittsburgh, Pittsburgh, PA 15260, USA}
\author{Peng-Cheng Lu}
\email{pclu@sdu.edu.cn}
\affiliation{School of Physics, Shandong University, Jinan, 250100, China}
\affiliation{Department of Physics and Astronomy, Michigan State University, East Lansing, MI 48824, USA}
\author{Tong Arthur Wu}
\email{tow39@pitt.edu}
\affiliation{Pittsburgh Particle Physics, Astrophysics, and Cosmology Center, \\
Department of Physics and Astronomy, University of Pittsburgh, Pittsburgh, PA 15260, USA}
{\hfill PITT-PACC-2510}
\begin{abstract}
We investigate the dimension-8 gluonic quartic gauge couplings (gQGCs) at future high-energy muon colliders through the process $\mu^{+}\mu^{-}\!\to gg\gamma$. Using detailed event simulation and optimized kinematic selections, we derive projected sensitivities to the Wilson coefficients and their associated new-physics scales, showing that muon colliders can probe deep into the multi-TeV regime and significantly surpass current LHC bounds. We further present the positivity bounds on those Wilson coefficients, as theoretical constraints from the fundamental principles of quantum field theory. Our results establish $\mu^{+}\mu^{-}\!\to gg\gamma$ as one of the most sensitive probes of dimension-8 new physics at future muon colliders.
\end{abstract}
\maketitle
\section{\label{sec1}Introduction}
% SM & NP
The standard model (SM) of particle physics remains the most successful theory describing the fundamental particles and their interactions, having been extensively tested in high energy collider experiments. However, it is widely accepted that the SM is only an effective theory, valid below a certain energy scale $\Lambda$, beyond which new physics is expected to emerge. This expectation is supported by several tensions between theoretical predictions and experimental measurements, such as the neutrino mass~\cite{Farzan:2017xzy,Arguelles:2022tki} and other precision anomalies~\cite{Muong-2:2023cdq,Crivellin:2023zui}. 
Future colliders with higher energies and luminosities will push the precision frontier and may reveal subtle deviations from the SM predictions.

% SMEFT
Among the theoretical frameworks used to parameterize new-physics effects, the Standard Model Effective Field Theory (SMEFT)~\cite{Buchmuller:1985jz} is particularly powerful because of its model-independent structure.  In SMEFT, heavy degrees of freedom are integrated out, resulting in an infinite tower of higher-dimensional operators that are suppressed by powers of the cutoff scale $\Lambda$. The leading correction appears at dimension-5 via the Weinberg operator, which generates Majorana masses for neutrinos~\cite{Weinberg:1979sa}. Anomalous gauge couplings first appear at dimension-6, and have been extensively studied in both current and projected collider analyses~\cite{LHCHiggsCrossSectionWorkingGroup:2016ypw}. However, many couplings receive their leading new-physics effects only at dimension-8, motivating systematic studies of the complete set of dimension-8 operators. 

% Dim-8
At dimension-8, the leading higher-order operators can be classified into several categories associated with anomalous gauge interactions, including anomalous quartic gauge couplings (aQGCs)~\cite{Eboli:2006wa,Eboli:2016kko}, neutral triple gauge couplings (nTGCs)~\cite{ntgc1, ntgc2, ntgc3, Rahaman:2018ujg, Senol:2018cks}, and the gluonic quartic gauge couplings (gQGCs)~\cite{Ellis:2018cos, Ellis:2021dfa}. 
The aQGCs and nTGCs parameterize possible deviations in the self-interactions among the electroweak ${\rm SU}(2)_L \times {\rm U}(1)_Y$ gauge bosons.
These anomalous couplings have been extensively investigated at the LHC and proposed 
future lepton colliders through processes, such as vector boson scattering (VBS)~\cite{Guo:2019agy, Guo:2020lim, Jiang:2021ytz, Yang:2021pcf, Yang:2021ukg, Ari:2021rmx, Yang:2022fhw, Yang:2023gos, Zhang:2023khv, Zhang:2024ebl, Chen:2025mxf}, di-boson productions~\cite{Ellis:2019zex, Ellis:2020ljj, Yang:2021kyy, Fu:2021mub, Ellis:2022zdw, Jahedi:2022duc, Jahedi:2023myu, Ellis:2023ucy, Guo:2024qyx, Liu:2024tcz, Xie:2025izk}, and tri-boson productions~\cite{Yang:2020rjt, Dong:2023nir, Zhang:2023yfg, Zhang:2023ykh}, providing stringent bounds on the corresponding dimension-8 Wilson coefficients. 
Measurements of these couplings not only test the nature of electroweak symmetry breaking but also provide sensitive probes of new physics (NP) beyond the SM.
While the electroweak aQGC and nTGC sectors have been systematically studied, the gluonic sector, which is described by dimension-8 operators involving both QCD and electroweak field strengths, remains largely unexplored. 

% gQGC
A particularly interesting class of quartic gauge couplings, those between the gluons and electroweak gauge bosons, is absent in the SM but allowed in the SMEFT at dimension-8. 
The gQGC operators induce processes such as $gg\to\gamma\gamma$~\cite{Ellis:2018cos} and $gg\to Z\gamma$~\cite{Ellis:2021dfa}, which are loop-induced in the SM but appear at tree level in the presence of these operators, allowing clean theoretical interpretations with negligible interference from SM backgrounds. 
Furthermore, there is no interference from dimension-6 operators in these processes, making them ideal probes of purely dimension-8 effects. 

% BI-model and M-theory
The quartic couplings of gluons to electroweak gauge bosons are well-motivated in ultraviolet (UV) complete frameworks such as the  Born–Infeld (BI) extension of the SM, where nonlinearities in the gauge sector naturally generate gQGC terms governed by a single energy scale~\cite{Born:1934gh}. 
Such a BI-like structure also appears in string-inspired UV completions~\cite{Fradkin:1985qd, Callan:1997kz,Arunasalam:2017eyu}. 

% Positivity bounds
Beyond specific models, general constraints on dimension-8 operators can also be derived from the fundamental principles of quantum field theory via the so-called positivity bounds~\cite{Adams:2006sv,Distler:2006if,deRham:2017avq,deRham:2017zjm,Zhang:2018shp,Arkani-Hamed:2020blm,Bellazzini:2020cot,Yamashita:2020gtt,Li:2022rag, Bi:2019phv, Zhang:2020jyn, Gu:2020ldn,Chen:2023bhu}.  
These bounds ensure that the low-energy Wilson coefficients correspond to a causal, local, and unitary UV completion. 
Collider tests of positivity bounds have become an emerging direction for both aQGC and nTGC operators, yet their application to the gluonic sector remains unexplored.  
The gQGC processes, being purely dimension-8 and free from SM contamination, offer a uniquely clean laboratory to test positivity bounds in collider environments.

% Study at MuC
Previous studies at the LHC, such as $gg\!\to\!\gamma\gamma$~\cite{Ellis:2018cos} and $gg\!\to\!Z\gamma$~\cite{Ellis:2021dfa}, have constrained the gQGC scales to the order of a few~TeV.  
Future lepton colliders, characterized by high energy, high luminosity, and clean experimental conditions, such as the proposed muon collider (MuC), provide an ideal environment to probe such purely dimension-8 effects. 
In this work, we focus on the process $\mu^+\mu^- \to jj\gamma$, where the jets originate from gluons.  
This channel receives contributions from tri-boson production mediated by gQGC operators, allowing a direct probe of gluonic quartic gauge interactions at lepton colliders.  
We perform a dedicated cut-based and significance analysis at three benchmark center-of-mass (c.m.) energies of future MuC.  
Furthermore, we investigate the corresponding positivity bounds on the gQGC operator coefficients, providing theoretical consistency checks for the dimension-8 SMEFT framework. 
In this work, we present the first study of positivity bounds in the gluonic sector, combining collider-level sensitivity analysis with theoretical consistency checks based on these bounds.
%Ongoing investigations are centered on a 10 TeV with a targeted integrated luminosity of 10 $ab^{-1}$ \cite{Black:2022cth,Accettura:2023ked}. 

% Paper Structure
The remainder of this paper is organized as follows. 
In Sec.~\ref{sec2:gqgc}, we review the dimension-8 operators contributing to gQGCs. 
Sec.~\ref{sec:analysis} presents the kinematic features of the $\mu^+\mu^- \!\to gg\gamma$ process, and a cut-based analysis for signal and background selection, describing the kinematic variables, event selection criteria, and the Monte Carlo simulation setup. 
We evaluate the statistical significance and derive the expected constraints on the individual gQGC operator coefficients at various benchmark c.m.~energies of MuC. 
In Sec.~\ref{sec:positivity}, we discuss the theoretical consistency of these results in light of positivity bounds, deriving the corresponding inequalities among Wilson coefficients and their implications for UV completions. 
Finally, Sec.~\ref{sec:summary} summarizes the main results and provides an outlook for future 
studies.

%%%%%%%%%%%%%%%%%%%%%%%%%%%%%%%%%%%%%
\section{\label{sec2:gqgc}Gluonic QGC Operators and \texorpdfstring{$gg\gamma$}{gga} production}
%%%%%%%%%%%%%%%%%%%%%%%%%%%%%%%%%%%%%
Deviations in the self-interactions of gauge bosons from their SM predictions are among the most sensitive probes of NP beyond the SM.
A systematic and gauge-invariant framework to parameterize such effects is provided by the SMEFT~\cite{Buchmuller:1985jz}, which incorporates all possible higher-dimensional operators constructed from SM fields. 
The SMEFT respects the full SM gauge group ${\rm SU}(3)_c \times {\rm SU}(2)_L \times {\rm U}(1)_Y$, and organizes the higher-order interactions as an expansion in inverse powers of a heavy mass scale $M$ that characterizes the energy scale of new physics.

In this framework, the gQGCs first appear at the dimension-8 level, suppressed by $1/M^4$~\cite{Ellis:2018cos}:
\begin{equation}
\begin{split}
\mathcal{L}_{gT} = \sum_{i=0}^{7} \frac{1}{16 M_i^4} \mathcal{O}_{gT,i}.
\end{split}
\label{eq.LgTi07M}
\end{equation}
For convenience, we define $f_i \equiv 1/M_i^4$, so that Eq.~(\ref{eq.LgTi07M}) can be rewritten as
\begin{equation}
\begin{split}
\mathcal{L}_{gT} = \sum_{i=0}^{7} \frac{f_i}{16} ~\mathcal{O}_{gT,i}.
\end{split}
\label{eq.LgTi07fi}
\end{equation}

The relevant gQGC operators are,
\begin{subequations}
\label{basis}
\begin{eqnarray}
  \mathcal{O}_{gT,0}
& \equiv &
  \frac 1 {16 M^4_0}
       \sum_a G^a_{\mu \nu} G^{a, \mu \nu}
\times \sum_i W^i_{\alpha \beta} W^{i, \alpha \beta} \,,
\\
  \mathcal{O}_{gT,1}
& \equiv &
  \frac 1 {16 M^4_1}
       \sum_a G^a_{\alpha \nu} G^{a, \mu \beta}
\times \sum_i W^i_{\mu \beta} W^{i, \alpha \nu} \,,
\\
  \mathcal{O}_{gT,2}
& \equiv &
  \frac 1 {16 M^4_2}
       \sum_a G^a_{\alpha \mu} G^{a, \mu \beta}
\times \sum_i W^i_{\nu \beta} W^{i, \alpha \nu} \,,
\\
  \mathcal{O}_{gT,3}
& \equiv &
  \frac 1 {16 M^4_3}
       \sum_a G^a_{\alpha \mu} G^a_{\beta \nu}
\times \sum_i W^{i, \mu \beta} W^{i, \nu \alpha} \,,
\\
  \mathcal{O}_{gT,4}
& \equiv &
  \frac 1 {16 M^4_4}
       \sum_a G^a_{\mu \nu} G^{a, \mu \nu}
\times B_{\alpha \beta} B^{\alpha \beta} \,,
\\
  \mathcal{O}_{gT,5}
& \equiv &
  \frac 1 {16 M^4_5}
       \sum_a G^a_{\alpha \nu} G^{a, \mu \beta}
\times B_{\mu \beta} B^{\alpha \nu} \,,
\\
  \mathcal{O}_{gT,6}
& \equiv &
  \frac 1 {16 M^4_6}
       \sum_a G^a_{\alpha \mu} G^{a, \mu \beta}
\times B_{\nu \beta} B^{\alpha \nu} \,,
\\
  \mathcal{O}_{gT,7}
& \equiv &
  \frac 1 {16 M^4_7}
       \sum_a G^a_{\alpha \mu} G^a_{\beta \nu}
\times B^{\mu \beta} B^{\nu \alpha} \,.
\end{eqnarray}
\end{subequations}
Here $G^a_{\mu \nu}$ denotes the gluon field strength, $W^i_{\mu \nu}$ and $B_{\mu \nu}$ are the electroweak field strengths, and the indices $i$ and $a$ are contracted.
Lorentz invariance further restricts the independent tensor contractions to four distinct structures, resulting in eight independent gQGC operators in total.
They can be grouped into four pairs,
\{$ \mathcal{O}_{gT,(0,4)}$,
$ \mathcal{O}_{gT,(1,5)}$,
$ \mathcal{O}_{gT,(2,6)}$,
$ \mathcal{O}_{gT,(3,7)}$\},
each pair sharing the same Lorentz structure.
The total and differential cross sections for the process $\mu^+\mu^-\!\to gg\gamma$ are largely determined by these Lorentz structures.

The dimension-8 gQGC operators provide an effective description of possible new interactions between gluons and electroweak gauge bosons. 
Their coefficients can be matched to various ultraviolet (UV) scenarios.
Among them, the BI extension of the SM is particularly motivated, since it naturally arises in string- and M-theory frameworks where gauge fields reside on D-branes. 
The BI Lagrangian takes the nonlinear form~\cite{Born:1934gh}
\begin{equation}
{\cal L}_{\rm BI}= \beta^2 \!\left[ 1- \sqrt{1 + \sum_{\lambda = 1}^{12} \frac {F^\lambda_{\mu \nu} F^{\lambda, \mu \nu}}{2 \beta^2} - \left( \sum_{\lambda = 1}^{12} \frac {F^\lambda_{\mu \nu} \widetilde F^{\lambda, \mu \nu}}{4 \beta^2} \! \right)^2} \;\right],
\label{eq:BI}
\end{equation}
where $\beta \equiv M^2$ is the BI nonlinearity scale and $\lambda$ runs over the 12 generators of the ${\rm SU}(3)_c \times {\rm SU}(2)_L \times {\rm U}(1)_Y$ gauge group. 
Originally introduced to impose an upper bound on field strengths, the BI theory has been extensively explored in nonlinear electrodynamics and appears naturally in D-brane effective actions~\cite{Fradkin:1985qd, Callan:1997kz}.

Expanding the BI action in powers of $1/M^4$ generates correlated quartic self-interactions among all SM gauge fields, corresponding to a specific linear combination of the eight gQGC operators introduced in~\cite{Ellis:2018cos}:
\begin{equation}
{\cal O}_{\rm BI} \propto 
\mathcal{O}_{gT,0}-2\mathcal{O}_{gT,1}+4\mathcal{O}_{gT,3}+\mathcal{O}_{gT,4}-2\mathcal{O}_{gT,5}+4\mathcal{O}_{gT,7},
\label{eq:BI_O}
\end{equation}
where all operators share the common cutoff scale $M$ and fully correlated coefficients~\cite{Ellis:2021dfa}. 
This correspondence provides a theoretically motivated benchmark for interpreting collider measurements in terms of UV-complete nonlinear gauge dynamics.

Using the ATLAS 13~TeV diphoton data, Ref.~\cite{Ellis:2018cos} derived the 95\%~C.L. lower bounds on the effective operator scales:
\begin{equation}
\begin{split}
M_0\geq 1040~{\rm GeV},\;&\quad M_1\geq 777~{\rm GeV},\\
M_2\geq 750~{\rm GeV},\;&\quad M_3\geq 709~{\rm GeV},\\
M_4\geq 1399~{\rm GeV},\;&\quad M_5\geq 1046~{\rm GeV},\\
M_6\geq 1010~{\rm GeV},\;&\quad M_7\geq 954~{\rm GeV}.
\end{split}
\label{eq:gQGC_limit}
\end{equation}
These limits correspond to an effective BI scale $M_{\rm BI}\!\sim\!1~{\rm TeV}$. 
Further analyses combining $gg\!\to\!Z\gamma$ and $\gamma\gamma$ channels have extended the sensitivity to $M_i \!\sim\!2~{\rm TeV}$~\cite{Ellis:2021dfa}. 
For comparison, the ATLAS measurement of light-by-light scattering in heavy-ion collisions excluded the pure QED BI scale below $\sim\!100~{\rm GeV}$~\cite{bi3}, while studies at future MuC indicate that BI scales of several~TeV could be probed through VBS processes~\cite{Yang:2023gos, Zhang:2023khv}. 

%%%%%%%%%%%%%%%%%%%%%%%%%%%%%%%%%%%%%
\section{\label{sec:analysis} Probing gQGCs with \texorpdfstring{$jj\gamma$}{jja} Production at MuC } 
%%%%%%%%%%%%%%%%%%%%%%%%%%%%%%%%%%%%%
%%%%%%%%%%%%%%%%%%%%%%%%%%%%%%%%%%%%%
\subsection{\label{sec:cross_section} Kinematic Features of the \texorpdfstring{$\mu^+\mu^-\to jj\gamma$}{jja} Process} 
%%%%%%%%%%%%%%%%%%%%%%%%%%%%%%%%%%%%%
The BI-type extensions of the SM provide a theoretically well-motivated UV completion of the dimension-8 gQGC operators, in which all gauge sectors are correlated through a single nonlinearity scale. 
Collider measurements of $\mu^+\mu^-\!\to gg\gamma$ thus offer a powerful probe of such nonlinear gauge dynamics, potentially revealing or constraining new physics at the multi-TeV scale.

Each gauge field strength ($G^a_{\mu\nu}$, $W^i_{\mu\nu}$, and $B_{\mu\nu}$) contains one derivative and therefore contributes a factor of momentum $p$ to the scattering amplitude. 
The four field strengths in a dimension-8 operator thus lead to an amplitude scaling as $\mathcal{M}\!\propto\!p^4$. We neglect the interference between the gQGC amplitude and the loop-induced SM contribution, which is loop-suppressed compared to the squared gQGC amplitude. 
Including the flux factor, which scales as $1/s$ for high-energy scattering, the total cross section induced by the gQGC operators grows as $\sigma\!\propto\!s^3\!\sim\!p^6$.
Consequently, the signal cross sections associated with $\mathcal{O}_{gT,i}$ rise rapidly with increasing c.m. energy.
The different Lorentz structures in Eq.(\ref{basis}) result in distinct total cross sections for the process $\mu \mu \rightarrow gg\gamma$:
\begin{align}
    \sigma^{\phantom{a}}_{\mathcal{O}_{gT,i}} = \begin{dcases}
        \big(\frac{f_i}{16}\big)^2 \frac{\alpha\, s^3}{360\pi^2} & \quad i=0, \\[1mm]
        \big(\frac{f_i}{16}\big)^2 \frac{7\alpha\, s^3}{4320\pi^2} & \quad i=1, \\[1mm]
         \big(\frac{f_i}{16}\big)^2 \frac{\alpha\, s^3}{2880\pi^2} & \quad i=2, \\[1mm]
         \big(\frac{f_i}{16}\big)^2 \frac{\alpha\, s^3}{2160\pi^2} & \quad i=3, \\
         \, 5\, \sigma^{\phantom{a}}_{\mathcal{O}_{gT,i-4}} & \quad 4\leq i \leq 7 ,
    \end{dcases}
    \label{eq:sigma_gT}
\end{align}
where $\alpha$ is the fine-structure constant, $s$ is the squared c.m.~energy, and the $Z$ boson mass is neglected. 
% The $Z$ boson mass is neglected in the high energy limit. 
Cross sections for $\mathcal{O}_{gT,(4,5,6,7)}$ are five times larger than those for $\mathcal{O}_{gT,(0,1,2,3)}$.

At very high energies, one must in principle examine the validity of the SMEFT expansion.
In the present case, however, the amplitudes induced by the gQGC operators are suppressed by internal propagators, and the unitarity bounds become relevant only at extremely low luminosities~\cite{Yang:2020rjt, Fu:2021mub}.
Consequently, the process $\mu^+\mu^-\!\to gg\gamma$ remains safely within the unitarity bounds for the collider scenarios considered in this study.

\begin{figure}[H]
    \centering
    \includegraphics[width=0.4\linewidth]{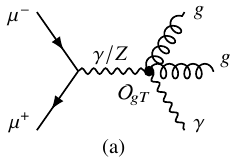}
    \includegraphics[width=0.8\linewidth]{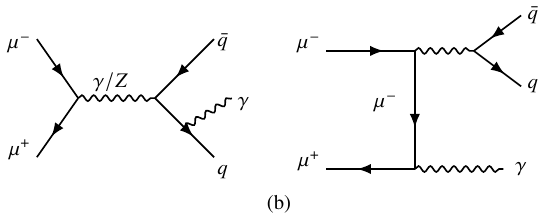}    
    \caption{Representative Feynman diagrams for (a) $\mu^+\mu^-\!\to gg\gamma$ induced by gQGC operators, and (b) the dominant SM background $\mu^+\mu^-\!\to q \bar q \gamma$. }
    \label{fig:feynman-NP}
\end{figure}
In the SM, there is no tree-level process corresponding to $\mu^+\mu^-\!\to gg\gamma$.
However, at a collider, quark- and gluon-initiated jets cannot be perfectly distinguished. 
Therefore, the dominant SM background arises from $\mu^+\mu^-\!\to q\bar{q}\gamma$.
Since these final states are different, no tree-level interference occurs between the SM and the gQGC induced amplitudes.
Representative Feynman diagrams for the signal and the SM background are shown in Fig.~\ref{fig:feynman-NP}. 
The SM background consists mainly of two components:
an $s$-channel process mediated by $\gamma/Z$ exchange
and a $t$-channel diboson process. 
The $s$-channel contribution exhibits a resonant enhancement near the
$Z$ pole ($\sqrt{s}\!\simeq\! m_Z$) and decreases as $1/s$ at higher energies. 
The $t$-channel amplitude is enhanced in the forward region as the momentum transfer $t\!\to\!0$, producing a pronounced forward peak in the photon and jet angular distributions. 
As the collider energy increases, the $t$-channel events become more forward-collimated, while the observable cross section within the detector acceptance decreases more rapidly than the simple $1/s$ behavior of the $s$-channel process. 
The total SM background cross sections at three benchmark
c.m. energies of the muon collider are summarized in
Table~\ref{table:XBG}. 
They decrease with energy, from about $20~\mathrm{pb}$ at $\sqrt{s}=3~\mathrm{TeV}$ to about $0.15~\mathrm{pb}$ at $\sqrt{s}=30~\mathrm{TeV}$.

\begin{table}[htbp]
\centering
\caption{\label{table:XBG} The cross section of the SM backgrounds.}
\vspace{2pt}
\setlength{\tabcolsep}{14pt}
\begin{tabular}{c|c|c|c}
\hline
 $\sqrt{s}\;({\rm TeV})$  & 3 & 10 & 30 \\
\hline 
 $\sigma _{\rm SM}\;({\rm pb})$ & $20.0$ &  $1.24$ & $0.153$ \\
\hline
\end{tabular}
\end{table}
%%%%%%%%%%%%%%%%%%%%%%%%%%%%%%%%%%%%%
\subsection{\label{sec:signal&background} Cut-based analysis for \texorpdfstring{$jj\gamma$}{jja} signal} 
%%%%%%%%%%%%%%%%%%%%%%%%%%%%%%%%%%%%%
In this section, we develop an optimized kinematic selection strategy to identify gQGC induced signal events while suppressing the SM backgrounds. To investigate the collider signatures of gQGCs, we perform a detailed cut-based analysis of the process $\mu^+\mu^-\!\to gg\gamma$, where the two gluons manifest as reconstructed jets in the detector. The signal and background events are generated using \verb"MadGraph5_aMC@NLO"~\cite{Alwall:2014hca} with the gQGC operators implemented through a UFO model created using \verb"Feynrules"~\cite{Christensen:2008py}. Parton showering and hadronization are handled by \verb"PYTHIA8"~\cite{Sjostrand:2014zea}, followed by a fast detector simulation with \verb"DELPHES" employing the muon collider configuration card~\cite{deFavereau:2013fsa}. The subsequent kinematic reconstruction and analysis are performed using the \verb"MLAnalysis"~\cite{Guo:2023nfu}. 
\begin{figure*}[htbp]
\centering
\includegraphics[width=0.3\hsize]{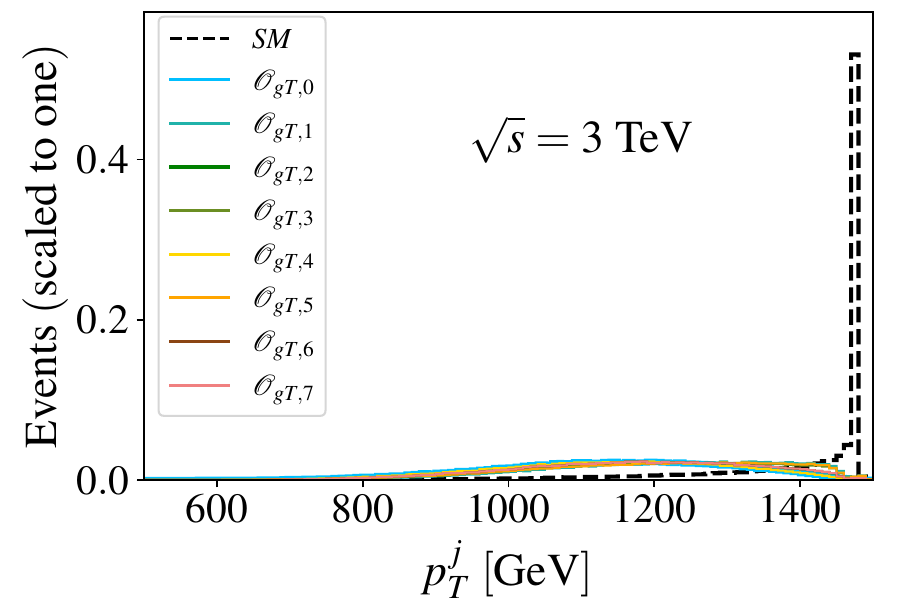}\quad
\includegraphics[width=0.3\hsize]{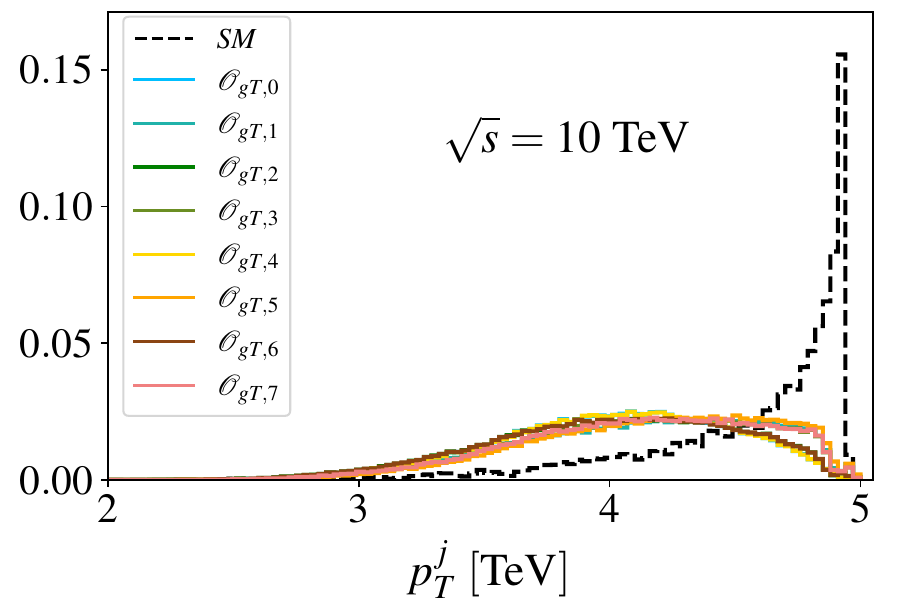}\quad
\includegraphics[width=0.3\hsize]{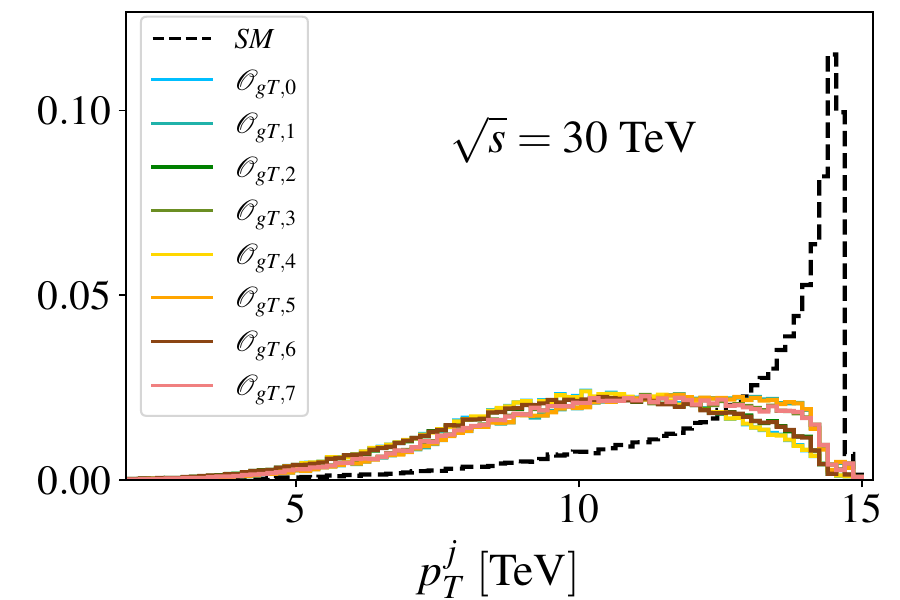}
\caption{\label{jpt}Normalized transverse momentum of leading jet ~$p_T^{j}$ distributions for signals generated by different operators and SM background at 3, 10 and 30 TeV~MuC.}
\end{figure*}
\begin{figure*}[htbp]
\centering
\includegraphics[width=0.3\hsize]{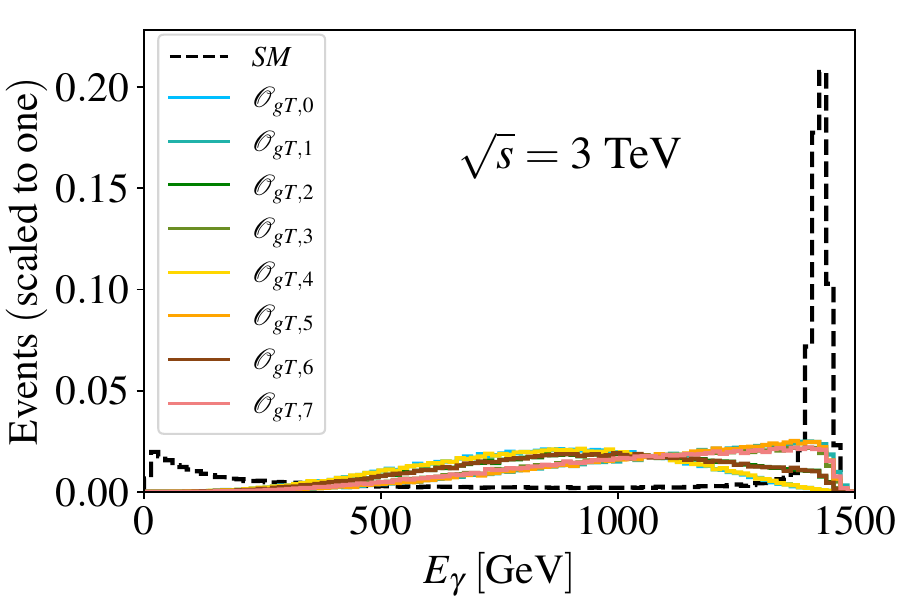}\quad
\includegraphics[width=0.3\hsize]{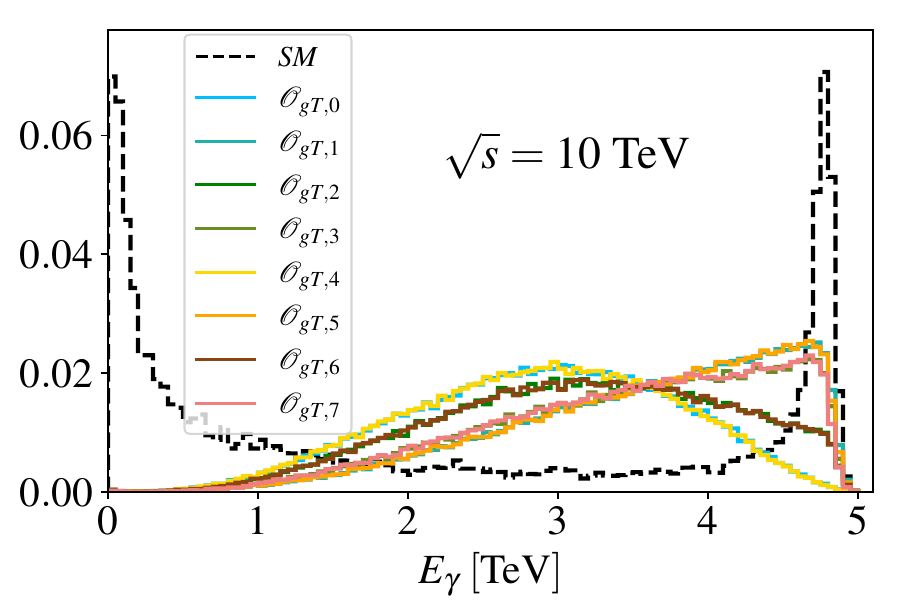}\quad
\includegraphics[width=0.3\hsize]{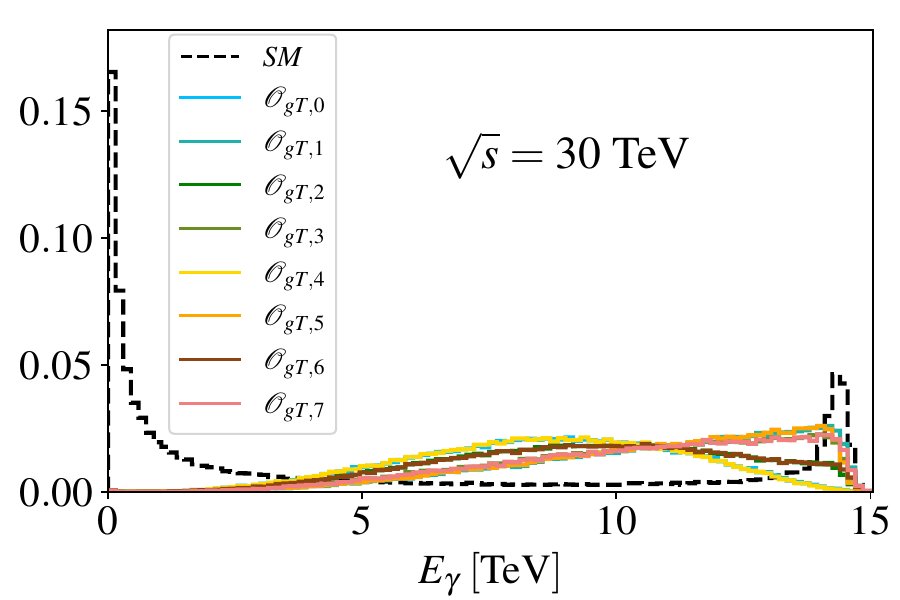}
\caption{\label{aE} The same configuration as in Fig.~\ref{jpt}, but for photon energy $E_{\gamma}$.}
\end{figure*}
\begin{figure*}[htbp]
\centering
\includegraphics[width=0.3\linewidth]{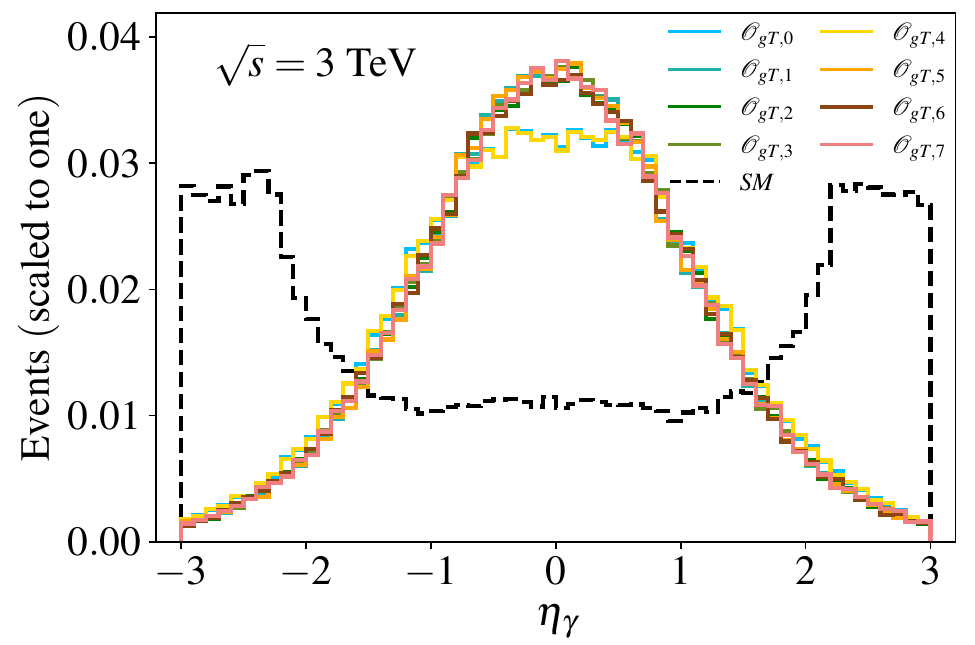}\quad
\includegraphics[width=0.3\linewidth]{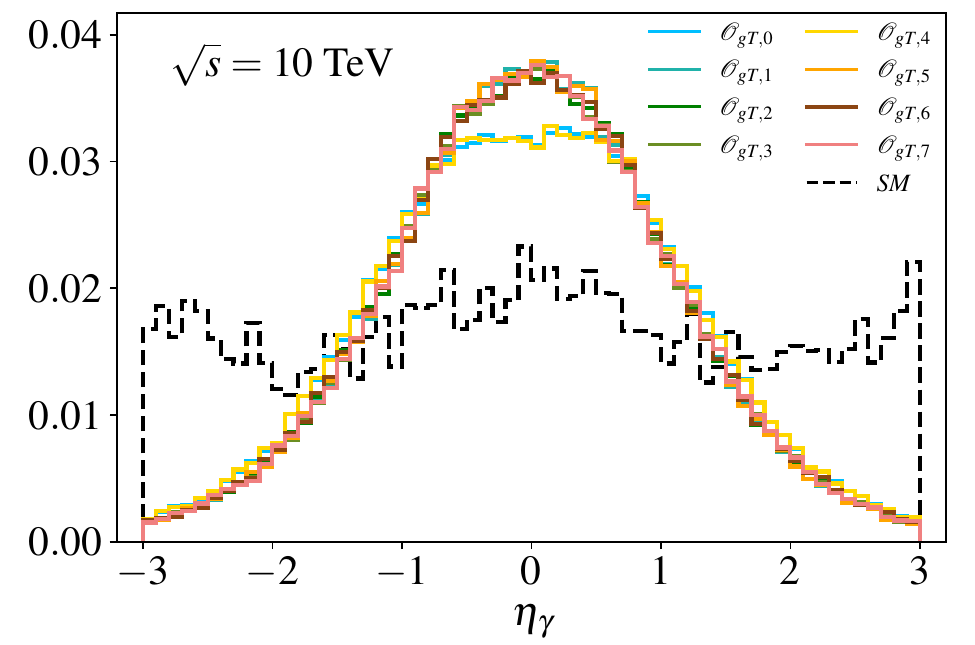}\quad
\includegraphics[width=0.3\linewidth]{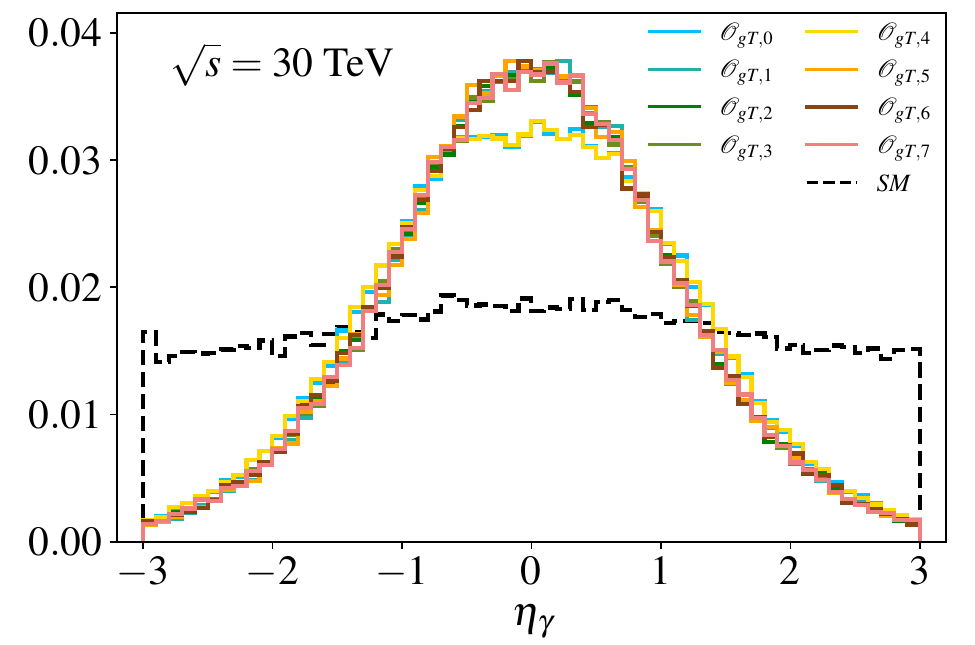}
\caption{\label{aETA}The same configuration as in Fig.~\ref{jpt}, but for pseudorapidity of photon  $\eta_{\gamma}$.}
\end{figure*}
\begin{figure*}[htbp]
\centering
\includegraphics[width=0.3\hsize]{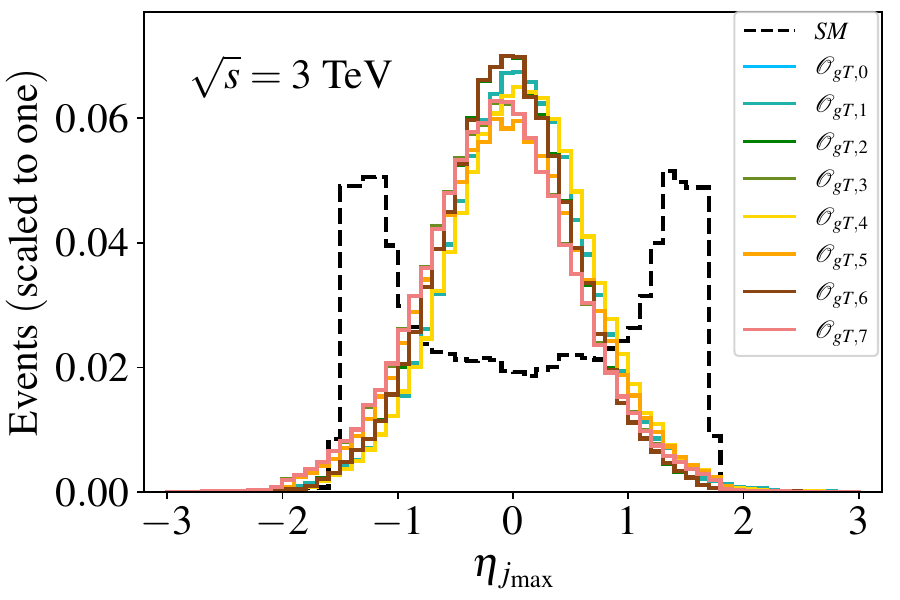}\quad
\includegraphics[width=0.3\hsize]{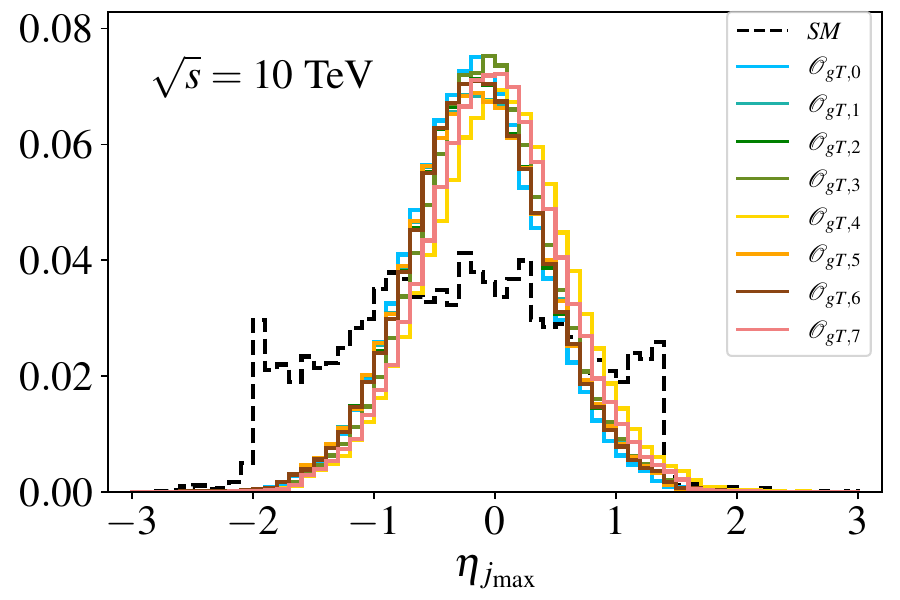}\quad
\includegraphics[width=0.3\hsize]{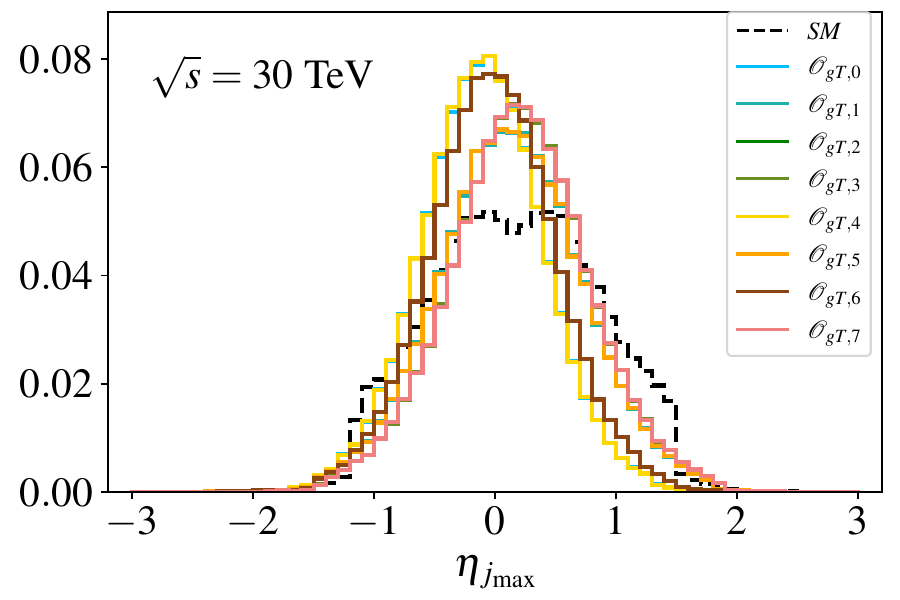}
\caption{\label{jETA}The same configuration as in Fig.~\ref{jpt}, but for pseudorapidity of leading jet $\eta_{j_{max}}$.} 
\end{figure*}
\begin{figure*}[htbp]
\centering
\includegraphics[width=0.3\hsize]{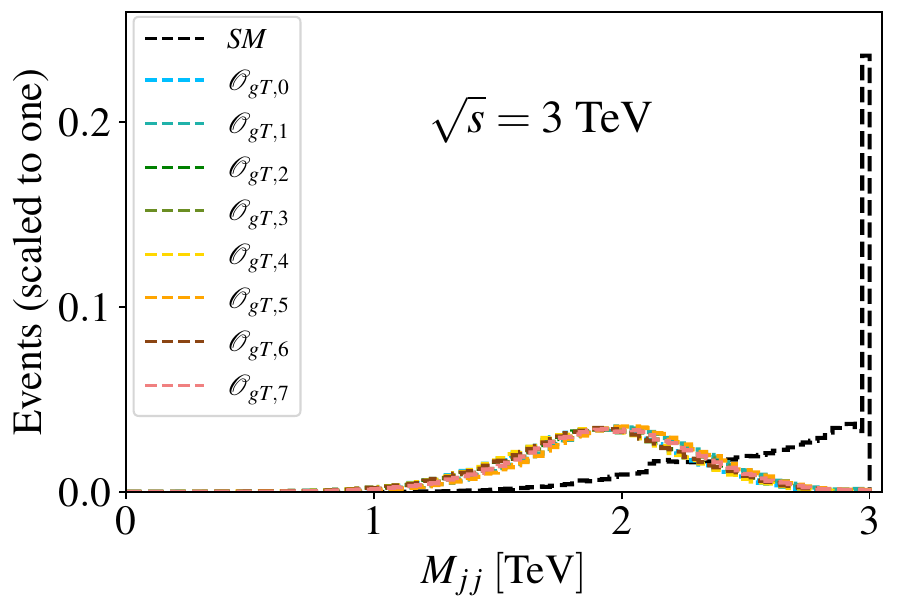}\quad
\includegraphics[width=0.3\hsize]{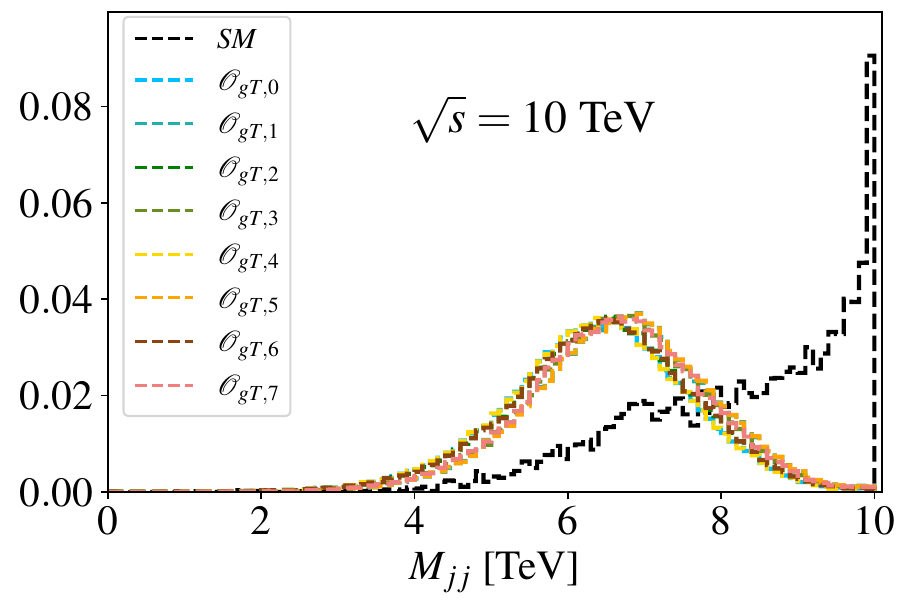}\quad
\includegraphics[width=0.3\hsize]{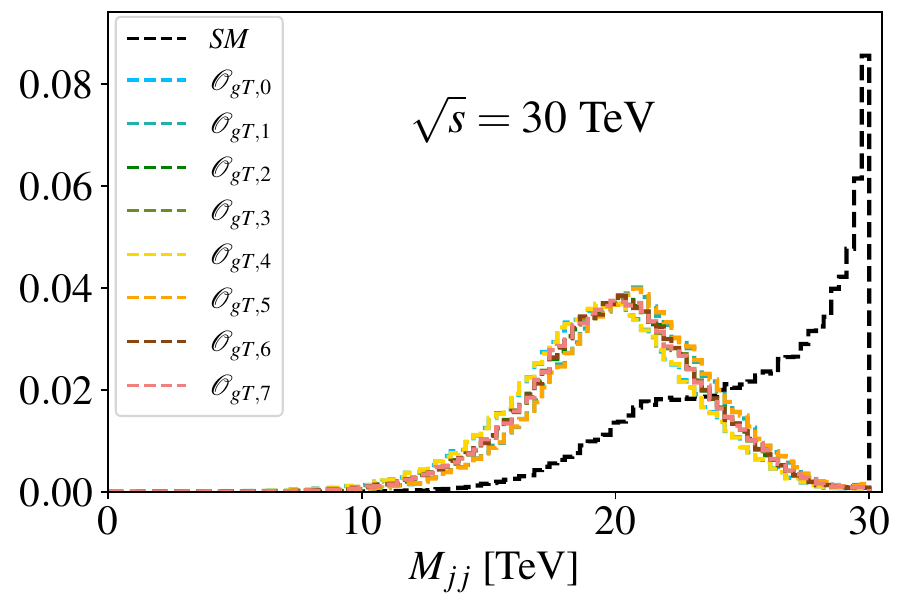}
\caption{\label{mjj}The same configuration as in Fig.~\ref{jpt}, but for the reconstructed invariant mass of di-jets  $M_{jj}$.}
\end{figure*}

In all analyses, we follow the conventional single-operator approach and activate one operator $\mathcal{O}_{gT,i}$ at a time. At a muon collider, the gQGC operators contribute to this process through an effective $s$-channel exchange of a virtual $\gamma/Z$ exchange, as illustrated in Fig.~\ref{fig:feynman-NP}(a). The resulting topology corresponds to a triboson configuration and leads to two energetic jets and one energetic photon in the final state.  

The SM background receives contributions mainly from two topologies: an $s$-channel $\gamma/Z$ exchange accompanied by QED radiation, and a $t$-channel diboson process mediated by a virtual boson. In the $s$-channel configuration, the photon is radiated either from the initial-state muon (ISR) or from one of the final-state quark lines (FSR). ISR favors soft or collinear photon emission, while FSR photons are primarily emitted collinearly with the quark due to the $1/t$ enhancement of the propagator. The $t$-channel topology exhibits a characteristic forward enhancement as $t\!\rightarrow\!0$, producing a forward-peaked photon and dijet system. By applying suitable $\eta_{j}$ and $\eta_{\gamma}$ cuts, the impact of this forward $t$-channel background can be effectively reduced. 

To ensure well-reconstructed final states, we impose a set of basic preselection requirements.  
Events are required to contain at least two reconstructed jets and at least one photon:
\begin{equation}
N_j \ge 2,\qquad N_\gamma \ge 1.
\end{equation}
The jets and photons must satisfy the transverse-momentum and pseudorapidity criteria:
\begin{align}
p_T^j &> 20~\mathrm{GeV},\quad |\eta_j| < 5,\\
p_T^\gamma &> 10~\mathrm{GeV},\quad |\eta_\gamma| < 2.5.
\end{align}
To ensure well-isolated final-state candidates and remove soft/collinear QED radiation, we further require that
\begin{equation}
\Delta R(\gamma,j) > 0.4,\qquad 
\Delta R(j,j) > 0.4,
\end{equation}
where particle flow isolation $\Delta R=\sqrt{(\Delta\eta)^2+(\Delta\phi)^2}$ denotes the separation in the pseudorapidity–azimuth plane. These preselection cuts efficiently remove soft or collinear ISR and FSR photons, and suppress highly forward events.

As shown in Eq.~\eqref{eq:sigma_gT}, the total cross section of gQGC induced process scales as \(\sigma\propto s^3\). Thus, gQGC signal tends to populate the hard-scattering region, leading to broad and extended distributions in the transverse momentum, energy, and angular variables of both the jets and the photon. In contrast, the SM background $\mu^+\mu^-\!\to q\bar q \gamma$ exhibits characteristic features of QED radiation and propagator-enhanced forward scattering. In the $s$-channel contribution, ISR photons generate a soft or collinear peak, while FSR photons are preferentially emitted along the quark direction due to the quark-propagator enhancement. The $t$-channel component produces a sharply forward-peaked photon and dijet system, though most of these events fall outside the detector acceptance after the preselection cuts. Consequently, the visible SM background is dominated by the $s$-channel, with the hard ISR peak near $E_\gamma\simeq\sqrt{s}/2$ and a characteristic soft-photon enhancement at low energy. 

Figure~\ref{jpt} shows the normalized transverse-momentum distribution of the leading jet for both signal and background processes at $\sqrt{s}=3,\,10,$ and $30~\mathrm{TeV}$. SM background events exhibit a strong preference for $p_T^j\!\sim\!\sqrt{s}/2$, reflecting the back-to-back two-body kinematics of $s$-channel production. In contrast, the gQGC signals show much broader $p_T^j$ spectra, extending well beyond the SM peak.

The photon energy distributions provide additional discrimination. 
As shown in Fig.~\ref{aE}, SM events display two characteristic enhancements: 
a soft peak from ISR and a hard peak near $E_\gamma\!\simeq\!\sqrt{s}/2$. Between two sharp peaks, the spectrum dips due to the suppression of intermediate ISR energies. The gQGC signals predominantly occupy this intermediate region, producing broader distributions. 
The $E_\gamma$ distributions of the four operators that form two Lorentz pairs, $\{1,5\}$ and $\{3,7\}$, exhibit a peak shift toward $E_\gamma\simeq\sqrt{s}/2$.
This arises from their common Lorentz structures, whose field-strength contractions enhance amplitude terms proportional to $(p_\gamma\!\cdot p_g)^2$, thereby favoring kinematic configurations in which the photon carries a large fraction of the total energy.  
The same hard photon enhanced behavior appears for the SU(2)$_L$ and U(1)$_Y$ types of these operators because they share identical Lorentz contractions.

Angular observables also carry substantial discriminating power. Figures~\ref{aETA} and \ref{jETA} display the pseudorapidity distributions of the photon and leading jet. The SM background, dominated by FSR and forward $t$-channel topologies, peaks at large $|\eta|$, whereas the gQGC-induced photons and jets remain central. 
We therefore apply $\eta_\gamma$ and $\eta_j$ selections that retain the signal while efficiently rejecting forward SM events. 

Finally, the dijet invariant-mass distribution $M_{jj}$ provides complementary separation. As shown in Fig.~\ref{mjj}, the SM background peaks near collision energy, consistent with the limited available energy after ISR emission. The signal, however, tends to cluster around $M_{jj}\!\sim\!(2/3)\sqrt{s}$, reflecting the harder three-body kinematics of gQGC production. An upper bound on $M_{jj}$ effectively suppresses the SM background with minimal signal loss. The combined use of transverse-momentum, energy, angular, and invariant-mass observables yields a highly efficient separation between gQGC signals and SM backgrounds.  
The complete set of selection requirements applied at each collider energy is summarized in Table~\ref{table:cuts}. 
The cross sections and selection efficiencies for both the SM background and signals after applying optimized kinematic selection strategy are summarized in Table~\ref{tab:cut-efficiency}. 
\begin{table}[htbp]
\centering
\caption{\label{table:cuts}Summary of the optimized event selection strategies used for $\mu^{+}\mu^{-}\!\to jj\gamma$ at various c.m. energies.}
\setlength{\tabcolsep}{5pt}
\vspace{2pt}
\begin{tabular}{c|c|c|c|c|c}
\hline
 $\sqrt{s}$    &  $p_T^j$      &$E_\gamma$    & $|\eta_\gamma|$& $|\eta_{j_\text{max}}|$& $M_{jj}$ \\
 $({\rm TeV})$ &  $({\rm TeV})$&$({\rm TeV})$ &              &             &$({\rm TeV})$\\
\hline 
 $3$           &  $<1.4$      & $(0.25,1.4)$ &              &              & $<2.45$  \\
 $10$          &  $<4.6$      & $(1.1,4.5)$  &   $<1.5$     &   $<1.5$     & $<8.30$ \\
 $30$          &  $<14.5$     & $(3.6,13.7)$ &              &              & $<24.5$  \\
\hline  
\end{tabular}
\end{table}
\begin{table*}[htbp]
\centering
\caption{Contributions of the SM and individual dimension-8 gQGC operators after sequential cuts at different $\sqrt{s}$. The last row reports the overall selection efficiency $\epsilon$ for each column.}
\vspace{2pt}
\label{tab:cut-efficiency}
\begin{tabular}{c|c|c|c|c|c|c|c|c|c|c}
\hline
$\sqrt{s}$ (TeV) & Cut & SM & $\mathcal{O}_{gT,0}$ & $\mathcal{O}_{gT,1}$ & $\mathcal{O}_{gT,2}$ & $\mathcal{O}_{gT,3}$ & $\mathcal{O}_{gT,4}$ & $\mathcal{O}_{gT,5}$ & $\mathcal{O}_{gT,6}$ & $\mathcal{O}_{gT,7}$ \\
 &  & (fb) & (fb) & (fb) & (fb) & (fb) & (fb) & (fb) & (fb) & (fb) \\
\hline
\multirow{6}{*}{3}
      & $p_T^j$            & 17.38 & 10.29 & 10.31 & 10.32 & 10.32 & 10.27 & 10.29 & 10.32 & 10.31 \\
      & $E_\gamma$         &  4.12 & 10.21 &  9.58 &  9.97 &  9.66 & 10.18 &  9.55 &  9.96 &  9.66 \\
      & $\eta_\gamma$      &  2.70 &  9.03 &  8.65 &  8.99 &  8.69 &  9.01 &  8.62 &  8.97 &  8.70 \\
      & $\eta_{j_{\max}}$  &  2.42 &  8.60 &  8.39 &  8.63 &  8.48 &  8.59 &  8.38 &  8.63 &  8.48 \\
      & $M_{jj}$           &  1.68 &  8.08 &  7.67 &  8.01 &  7.81 &  8.06 &  7.69 &  8.00 &  7.85 \\
      \cline{2-11}
      & $\epsilon$         &  8.4\% & 72.1\% & 68.5\% & 71.6\% & 69.7\% & 72.1\% & 68.7\% & 71.5\% & 70.0\% \\
\hline
\multirow{6}{*}{10}
      & $p_T^j$            & 1.06 & 0.83 & 0.84 & 0.83 & 0.83 & 0.83 & 0.82 & 0.83 & 0.83 \\
      & $E_\gamma$         & 0.29 & 0.81 & 0.65 & 0.75 & 0.67 & 0.81 & 0.64 & 0.75 & 0.66 \\
      & $\eta_\gamma$      & 0.19 & 0.72 & 0.59 & 0.67 & 0.60 & 0.71 & 0.58 & 0.67 & 0.60 \\
      & $\eta_{j_{\max}}$  & 0.17 & 0.68 & 0.56 & 0.64 & 0.58 & 0.68 & 0.56 & 0.64 & 0.58 \\
      & $M_{jj}$           & 0.12 & 0.64 & 0.53 & 0.61 & 0.56 & 0.64 & 0.53 & 0.61 & 0.56 \\
      \cline{2-11}
      & $\epsilon$         & 10.0\% & 70.6\% & 58.9\% & 66.9\% & 61.2\% & 70.6\% & 58.8\% & 66.8\% & 61.3\% \\
\hline
\multirow{6}{*}{30}
      & $p_T^j$            & 0.133 & 0.46 & 0.47 & 0.47 & 0.46 & 0.47 & 0.47 & 0.47 & 0.46 \\
      & $E_\gamma$         & 0.032 & 0.45 & 0.38 & 0.42 & 0.39 & 0.45 & 0.38 & 0.42 & 0.39 \\
      & $\eta_\gamma$      & 0.020 & 0.40 & 0.34 & 0.38 & 0.35 & 0.40 & 0.34 & 0.38 & 0.35 \\
      & $\eta_{j_{\max}}$  & 0.018 & 0.38 & 0.33 & 0.36 & 0.34 & 0.38 & 0.33 & 0.36 & 0.34 \\
      & $M_{jj}$           & 0.0137 & 0.35 & 0.30 & 0.33 & 0.31 & 0.35 & 0.30 & 0.33 & 0.31 \\
      \cline{2-11}
      & $\epsilon$         &  8.9\% & 68.8\% & 59.1\% & 65.7\% & 61.5\% & 68.8\% & 59.2\% & 65.7\% & 61.5\% \\
\hline
\end{tabular}
\end{table*}

%%%%%%%%%%%%%%%%%%%%%%%%%%%%%%%%%%%%%%%%%%%%%%%%%%%%%%%%%%%%%%%%%%%%%%%%%%%%%%%%%%%%%%%%
\subsection{\label{sec:significance} Prospective sensitivities at MuC and constraints on gQGC}
%%%%%%%%%%%%%%%%%%%%%%%%%%%%%%%%%%%%%%%%%%%%%%%%%%%%%%%%%%%%%%%%%%%%%%%%%%%%%%%%%%%%%%%%
In this section, we estimate the sensitivity of the process $\mu \mu \rightarrow gg\gamma$ to the dimension-8 gQGC operators and derive the corresponding expected bounds on their Wilson coefficients and cutoff scales. The statistical significance is defined as
\begin{align}
   \mathcal{S}_{\rm stat} = \frac{\sigma_{\rm gQGC} \times \epsilon_{\rm gQGC}}{\sqrt{\sigma_{\rm gQGC} \times \epsilon_{\rm gQGC} + \sigma_{\rm SM} \times \epsilon_{\rm SM}}} \times \sqrt{L},
\end{align}
where $\sigma_{\rm gQGC}$ and $\sigma_{\rm SM}$ are the signal and background cross sections, respectively. $\epsilon_{\rm gQGC}$ and $\epsilon_{\rm SM}$ are the corresponding selection efficiencies, and $L$ is the integrated luminosity. Three benchmark MuC configurations are considered: $\sqrt{s}=3~\mathrm{TeV}$ with $1~\mathrm{ab}^{-1}$, 
$\sqrt{s}=10~\mathrm{TeV}$ with $10~\mathrm{ab}^{-1}$, and $\sqrt{s}=30~\mathrm{TeV}$ with integrated luminosities of $10~\mathrm{ab}^{-1}$ and $90~\mathrm{ab}^{-1}$.

\begin{figure}
    \centering
    \includegraphics[width=1\hsize]{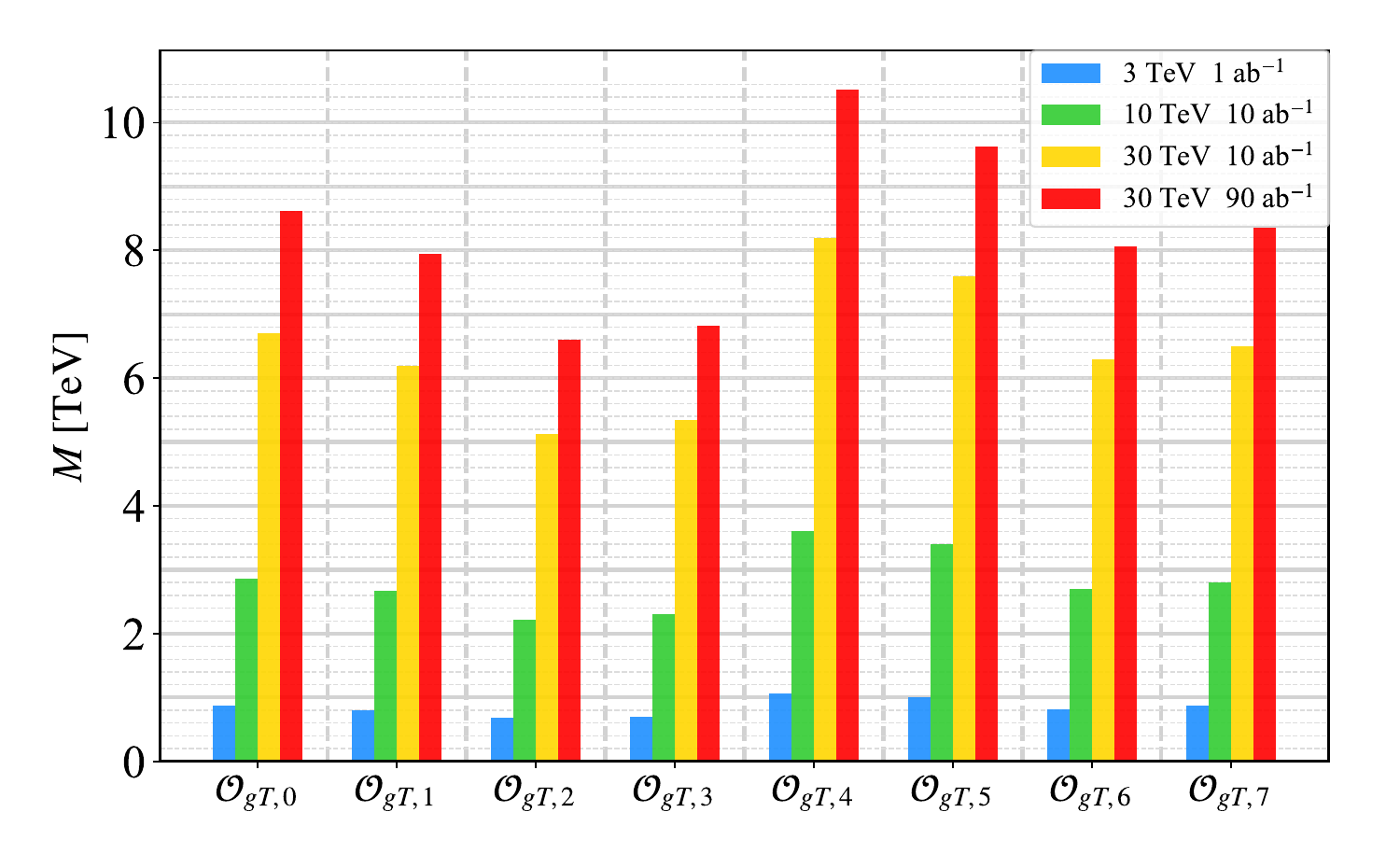}
    \caption{\label{M.pdf}The 95\% C.L. sensitivities on the cutoff scales $M_i$ of gQGC operators at MuC for the various benchmark configurations.}
    \label{fig:M}
\end{figure}

The projected sensitivities on the Wilson coefficients $f_i$ and the associated
new-physics scales $M_i = f_i^{-1/4}$ are summarized in 
Tables~\ref{tab.fi} and~\ref{tab.Mi}, while Fig.~\ref{fig:M} displays the corresponding 
reach on the cutoff scales.  
A clear hierarchy appears among the eight operators. Those involving the hypercharge field strength $B_{\mu\nu}$, $O_{gT,4}$ through $O_{gT,7}$,  lead to significantly larger cross sections due to the overall enhancement factor in Eq.~\eqref{eq:sigma_gT}. 
Consequently, they exhibit the best sensitivity across all collider energies.  
The operators $O_{gT,0}$ through $O_{gT,3}$ show a similar pattern within each Lorentz-structure pair, but with overall weaker reach.  
Since the signal cross section grows as $s^{3}$ while the SM background decreases approximately as $1/s$, the sensitivity improves rapidly with increasing collider energy.  
For instance, the reach on $M_0$ increases from about $0.9$~TeV at $\sqrt{s}=3~\mathrm{TeV}$ to roughly $2.8$~TeV at $10~\mathrm{TeV}$ and approaches $6$--$7$~TeV at $30~\mathrm{TeV}$ with $10~\mathrm{ab}^{-1}$.  
The improvement is even more striking for hypercharge-type operators $O_{gT,4}$–$O_{gT,7}$, where a $30~\mathrm{TeV}$ MuC with $90~\mathrm{ab}^{-1}$ can probe new physics scales in the $8$–$10$~TeV range.

\begin{table*}[htbp]
\begin{center}
\caption{\label{tab.fi}
Projected sensitivities at $2\sigma$, $3\sigma$, and $5\sigma$ on the gQGC coefficients (in units of ${\rm TeV}^{-4}$) at future MuC with various c.m. energies and integrated luminosities.}
%\scalebox{0.8}{
\begin{tabular}{c|c|c|c|c|c|c|c|c|c}
\hline
$\sqrt{s}$  & $\mathcal{S}_{stat}$ & $|f_0|$ & $|f_1|$ & $|f_2|$ & $|f_3|$ & $|f_4|$ & $|f_5|$ & $|f_6|$ & $|f_7|$ \\
\hline
$3$ TeV               & 2 & $<1.76$ & $<2.41$ & $<5.01$ & $<4.44$ & $<0.78$ & $<1.07$ & $<2.22$ & $<1.85$ \\
($1\,{\rm ab}^{-1}$)  & 3 & $<2.50$ & $<3.41$ & $<7.10$ & $<6.28$ & $<1.11$ & $<1.52$ & $<3.15$ & $<2.62$ \\
                      & 5 & $<3.97$ & $<5.32$ & $<11.2$ & $<9.86$ & $<1.77$ & $<2.37$ & $<5.03$ & $<4.40$ \\
\hline
$10$ TeV              & 2 & $<0.0150$ & $<0.0197$ & $<0.0420$ & $<0.0367$ & $<0.00667$ & $<0.0089$ & $<0.0188$ & $<0.0164$ \\
($10\,{\rm ab}^{-1}$) & 3 & $<0.0212$ & $<0.0281$ & $<0.0598$ & $<0.0522$ & $<0.00948$ & $<0.0126$ & $<0.0267$ & $<0.0233$ \\
                      & 5 & $<0.0341$ & $<0.0453$ & $<0.0963$ & $<0.0840$ & $<0.01520$ & $<0.0204$ & $<0.0431$ & $<0.0375$ \\
\hline
$30$ TeV             & 2 & $<0.000495$ & $<0.000682$ & $<0.00144$ & $<0.00125$ & $<0.000221$ & $<0.000303$ & $<0.000634$ & $<0.000561$ \\
($10\,{\rm ab}^{-1}$) & 3 & $<0.000753$ & $<0.00106$ & $<0.00217$ & $<0.00194$ & $<0.000336$ & $<0.000475$ & $<0.000973$ & $<0.000870$ \\
                      & 5 & $<0.00124$ & $<0.00176$ & $<0.00360$ & $<0.00322$ & $<0.000550$ & $<0.000788$ & $<0.00161$ & $<0.00144$ \\
\hline
$30$ TeV           & 2 & $<0.000182$ & $<0.000251$ & $<0.000530$ & $<0.000462$ & $<0.000081$ & $<0.000111$ & $<0.000233$ & $<0.000206$ \\
($90\,{\rm ab}^{-1}$) & 3 & $<0.000259$ & $<0.000356$ & $<0.000752$ & $<0.000656$ & $<0.000115$ & $<0.000158$ & $<0.000331$ & $<0.000293$ \\
                & 5 & $<0.000416$ & $<0.000573$ & $<0.001200$ & $<0.001050$ & $<0.000190$ & $<0.000254$ & $<0.000532$ & $<0.000471$ \\
\hline
\end{tabular}
%}
\end{center}
\end{table*}

\begin{table*}[htbp]
\begin{center}
\caption{\label{tab.Mi}Projected sensitivities for probing individual dimension-8 gQGC operators at $2\sigma$, $3\sigma$, and $5\sigma$ significance levels, expressed in terms of the corresponding cutoff scales $M_i$ (in units of TeV), at future MuC across different c.m. energies with integrated luminosities.
}
\begin{tabular}{c|c|c|c|c|c|c|c|c|c}
\hline
$\sqrt{s}$   & $\mathcal{S}_{\rm stat}$ & $M_0$ & $M_1$ & $M_2$ & $M_3$ & $M_4$ & $M_5$ & $M_6$ & $M_7$ \\
\hline
$3$ TeV                        & 2 & $>0.87$ & $>0.80$ & $>0.68$ & $>0.70$ & $>1.06$ & $>1.00$ & $>0.82$ & $>0.88$ \\
($1\,\mathrm{ab}^{-1}$)        & 3 & $>0.80$ & $>0.74$ & $>0.64$ & $>0.64$ & $>0.98$ & $>0.90$ & $>0.78$ & $>0.80$ \\
                               & 5 & $>0.71$ & $>0.66$ & $>0.56$ & $>0.58$ & $>0.88$ & $>0.82$ & $>0.68$ & $>0.70$ \\
\hline
$10$ TeV                        & 2 & $>2.86$ & $>2.67$ & $>2.22$ & $>2.30$ & $>3.60$ & $>3.40$ & $>2.70$ & $>2.80$ \\
($10\,\mathrm{ab}^{-1}$)        & 3 & $>2.62$ & $>2.44$ & $>2.04$ & $>2.10$ & $>3.24$ & $>3.00$ & $>2.48$ & $>2.56$ \\
                                & 5 & $>2.33$ & $>2.17$ & $>1.80$ & $>1.88$ & $>2.88$ & $>2.66$ & $>2.20$ & $>2.30$ \\
\hline
$30$ TeV  & 2 & $>6.70$         & $>6.19$ & $>5.12$ & $>5.34$ & $>8.20$ & $>7.60$ & $>6.30$ & $>6.50$ \\
($10\,\mathrm{ab}^{-1}$)        & 3 & $>6.04$ & $>5.54$ & $>4.64$ & $>4.78$ & $>7.40$ & $>6.82$ & $>5.68$ & $>5.84$ \\
                                & 5 & $>5.33$ & $>4.88$ & $>4.80$ & $>4.22$ & $>6.52$ & $>6.00$ & $>5.02$ & $>5.14$ \\
\hline
$30$ TeV                      & 2 & $>8.61$ & $>7.94$ & $>6.60$ & $>6.82$ & $>10.52$ & $>9.62$ & $>8.06$ & $>8.36$ \\
($90\,\mathrm{ab}^{-1}$)      & 3 & $>7.88$ & $>7.28$ & $>6.06$ & $>6.26$ & $>9.76$ & $>8.92$ & $>7.66$ & $>7.66$ \\
                              & 5 & $>7.00$ & $>6.46$ & $>5.40$ & $>5.50$ & $>8.60$ & $>8.00$ & $>6.82$ & $>6.82$ \\
\hline
\end{tabular}
\end{center}
\end{table*}

It is instructive to compare these results with existing LHC bounds.  
Analyse of projected sensitivity from $gg\to\gamma\gamma$ process at $\sqrt{s}=13$~TeV LHC with $3 \,\rm{ab}^{-1}$ luminosity have yielded limit of, for example, $M_0>2.1\,\rm{TeV}$~\cite{Ellis:2018cos}. While we find a $10\,\rm{TeV}$ MuC with $L=10 \,\rm{ab}^{-1}$ puts a stronger constraint of $M_0>2.6 \,\rm{TeV}$, and a $30\,\mathrm{TeV}$ machine reaches beyond $7$~TeV. 

These sensitivities carry important implications for UV scenarios.
In the BI extension of the Standard Model, the eight gQGC operators appear in a highly correlated linear combination governed by a single nonlinearity scale $M_{\rm BI}$.  
LHC heavy-ion light-by-light scattering currently constrains the hypercharge BI scale to roughly $100$~GeV, whereas our results demonstrate that muon colliders can probe BI scales at the multi-TeV level.  
A $10~\mathrm{TeV}$ muon collider is sensitive to $M_{\rm BI}\gtrsim 3~\mathrm{TeV}$, and a 
$30~\mathrm{TeV}$ machine with high luminosity can reach the $8$--$10$~TeV range, accessing the parameter region relevant for electroweak BI monopoles~\cite{Arunasalam:2017eyu} and many string-inspired UV completions.

%%%%%%%%%%%%%%%%%%%%%%%%%%%%%%%%%%%%%%%%%%%%%%%%%%%%%%%%%%%%%%%%%%%%%%%%%%
\section{Positivity bounds on tri-boson production}
\label{sec:positivity}
%%%%%%%%%%%%%%%%%%%%%%%%%%%%%%%%%%%%%%%%%%%%%%%%%%%%%%%%%%%%%%%%%%%%%%%%%%

Another interesting implication of the studies on these dimension-8 operators connects to the positivity bounds. Positivity bounds on the Wilson coefficients in the SMEFT originate from the requirement that any UV completion of the theory must respect the fundamental principles of quantum field theory, namely unitarity, locality, analyticity, and Lorentz invariance. 
These principles ensure that forward-limit scattering amplitudes are analytic functions of the Mandelstam variable $s$ and admit a twice-subtracted dispersion relation, from which the following inequality can be derived \cite{Adams:2006sv,Distler:2006if,deRham:2017avq}:
\begin{equation}
f_{\epsilon\Lambda}(s)
\equiv 
\frac{1}{2}\frac{\mathrm{d}^{2}B_{\epsilon\Lambda}(s)}{\mathrm{d}s^{2}} 
> 0 , 
\qquad 
|s|<(\epsilon\Lambda)^2 ,
\label{eq:positivity_general}
\end{equation}
where \(B_{\epsilon\Lambda}(s)\) denotes the amplitude with all poles and branch cuts below the subtraction scale \(\epsilon\Lambda\) subtracted out. 
The dispersion relation allows one to express $f_{\epsilon\Lambda}(s)$ as an integral of the discontinuity of the amplitude, and the optical theorem relates it to the total cross section, 
\(\mathrm{Disc}\,A(s)=2i\,\mathrm{Im}A(s)\propto\sigma_{\text{tot}}(s)>0\), which guarantees the positivity of \(f_{\epsilon\Lambda}(s)\).
Consequently, the coefficients of higher-dimensional operators in the EFT must satisfy a set of positivity inequalities if the underlying theory can be consistently UV completed \cite{Zhang:2018shp,Yamashita:2020gtt,Bi:2019phv,Zhang:2020jyn}.

In practice, such theoretical constraints are often difficult to probe experimentally, because they typically first appear at the level of dimension-8 or higher operators and can be obscured by lower-dimensional contributions. The gQGC operators, however, provide an ideal framework in which these conditions can be tested unambiguously, because there is no contribution from the SM or from any dimension-6 operator to the $gg\gamma\gamma$ vertex at tree level.

At the amplitude level, the forward scattering of vector bosons can be expressed as
\begin{equation}
A(s,t=0)
= A_{\text{SM}}(s)
+ \sum_i \frac{f_{i}}{16}\,\mathcal{M}_{i}(s) ,
\end{equation}
where \(\mathcal{M}_{i}(s)\) are the amplitudes from the dimension-8 gQGC operators \(\mathcal{O}_{gT,i}\).
Applying Eq.~(\ref{eq:positivity_general}) to this amplitude leads to following positivity bounds: 
\begin{align}
    f_2 &\geq 0, \\
    f_6 &\geq 0, \\
    4f_1+f_2&+2f_3 \geq 0, \\
    4f_5+f_6&+2f_7 \geq 0,
\end{align}
which indicate that only a restricted region of the full parameter space is compatible with a standard UV completion. These inequalities put additional constraints on the coefficients of gQGC operators, which are highly complementary to our results of the collider reach in Sec.~\ref{sec:significance}. Moreover, our analysis implies that positivity bounds can be potentially tested at MuC up to energy scales of a few TeV. 

Such measurement at multi-TeV MuC would therefore provide a unique opportunity to directly test whether the EFT parameter region inferred from collider data remains consistent with the fundamental principles of quantum field theory. Any significant violation of the positivity bounds would not only indicate the presence of new interactions beyond the SM, but would also challenge the validity of unitary and analytic quantum field theories at high energies.

%%%%%%%%%%%%%%%%%%%%%%%%%%%%%%%%%%%%%
\section{\label{sec:summary}Discussion and Conclusions} 
%%%%%%%%%%%%%%%%%%%%%%%%%%%%%%%%%%%%%

The process $\mu^+\mu^-\to gg\gamma$ provides one of the cleanest and most powerful probes of gluonic quartic gauge interactions, since no SM tree-level contribution exists and there is no interference from dimension-6 operators.  
The characteristic $s^{3}$ growth of the gQGC-induced amplitudes, together with the falling SM background, leads to rapidly improving sensitivity at higher collider energies.
The clean experimental environment, coupled with the ability to reach multi-TeV energies, significantly enhances sensitivity to dimension-8 operators encoding gQGC effects, positioning high-energy MuC as a powerful platform for advancing our understanding of the strong and electroweak sectors.

We have carried out a comprehensive investigation of the dimension-8 gQGCs at future MuC through the $jj\gamma$ signal. Through detailed signal and background analysis, our results demonstrate that a $3~\mathrm{TeV}$ MuC with $1~\mathrm{ab}^{-1}$ luminosity is already capable of probing new-physics scales at the TeV level; a $10~\mathrm{TeV}$ MuC with $10~\mathrm{ab}^{-1}$ extends the reach to the multi-TeV regime; and a $30~\mathrm{TeV}$ MuC with $10$--$90~\mathrm{ab}^{-1}$ can probe scales as high as $6$--$10~\mathrm{TeV}$ depending on the operator structure. These sensitivities represent a substantial improvement over current LHC bounds obtained from diphoton and $Z\gamma$ production \cite{Ellis:2018cos,Ellis:2021dfa}. 
The hierarchy of sensitivities among the operators originates from the fact that the hypercharge-type operators exhibit a stronger reach than the SU(2)$_L$-type ones. This is because the operators $\mathcal{O}_{gT,4}$–$\mathcal{O}_{gT,7}$ generate cross sections that are enhanced by an overall factor of $5$ relative to those induced by $\mathcal{O}_{gT,0}$–$\mathcal{O}_{gT,3}$. 
The $\mu^{+}\mu^{-}\!\to gg\gamma$ channel emerges as one of the most sensitive probes of anomalous gauge interactions in the gluonic sector, providing a uniquely clean avenue to explore dimension-8 new physics at future muon colliders. 

The sensitivity reach achieved in this study opens a direct window into a broad class of UV-motivated scenarios, including BI extensions of the SM, string-inspired nonlinear gauge dynamics, and other theories predicting dimension-8 gluonic interactions. 
BI extensions of the SM predict correlated quartic gauge interactions controlled by a single BI nonlinearity scale $M_{\rm BI}$, and our results show that muon colliders can probe the parameter region $M_{\rm BI}\sim 8$--$12~\mathrm{TeV}$, overlapping with the natural range expected in string-inspired BI dynamics.  

A further theoretical highlight of this study is the connection to positivity bounds.  
Since $\mu^{+}\mu^{-}\!\to gg\gamma$ receives contributions exclusively from dimension-8 operators, its inferred Wilson coefficients can be directly confronted with the positivity inequalities derived from unitarity, analyticity, and causality of the forward scattering amplitude.   
We find complementary constraints on the gQGC operators using positivity bounds. Moreover, our results implies that the projected sensitivities at a $10~\mathrm{TeV}$ MuC with $10~\mathrm{ab}^{-1}$ would be sufficient to test positivity at the TeV scale in the gluonic sector. 

In addition to the cut-based strategy used in this analysis, recent progress in machine-learning jet-substructure techniques has demonstrated remarkable improvements in quark--gluon discrimination.  
Methods such as convolutional neural networks applied to jet images, graph neural network architectures including ParticleNet \cite{Qu:2019gqs}, LundNet~\cite{Komiske:2018cqr}, and the jet-origin identification framework~\cite{Liang:2023wpt,Zhu:2025eoe}, have shown significant gains in tagging efficiency and background rejection relative to traditional observables. 
Future improvements using machine-learning-based jet tagging promise to further enhance the sensitivity, potentially enabling MuC to approach the theoretical limits of sensitivity to gQGCs. 
This motivates future studies incorporating advanced machine-learning algorithms into new physics search strategies at MuC. 

\vspace{1cm}
\section*{Acknowledgment}
\noindent We thank Tao Han, Ji-Chong Yang, Samuel Homiller and Man-Qi Ruan for the helpful discussion. This work was supported in part by the Natural Science Foundation of the Liaoning Scientific Committee No. JYTMS20231053, the National Natural Science Foundation of China under Grants No. 12405121, and the China Scholarship Council. 

\bibliographystyle{apsrev4-1}
\bibliography{gqgc-gga}

\end{document}